\documentclass{iopart}
\usepackage{epsfig}
\usepackage{subfigure}
\begin{document}
 
\article[Transverse hydrodynamics --- production of strangeness]
  {Strangeness in Quark Matter 2009, Buzios, Brazil}
  {Transverse hydrodynamics with sudden hadronization --- production of strangeness\footnote{Supported in part by the Polish Ministry of Science and Higher Education, grant N202 034 32/0918.}
}  
 
\author{W. Florkowski$^{1,2}$ and R. Ryblewski$^{2}$}

\address{$^1$ Institute of Physics, Jan~Kochanowski University, 25-406 Kielce, Poland}
\address{$^2$ The H.~Niewodnicza\'nski Institute of Nuclear Physics, 
Polish Academy of Sciences, 31-342 Krak\'ow, Poland} 
   
\begin{abstract}
We consider a physical scenario for ultra-relativistic heavy-ion collisions where, at the early stage, only transverse degrees of freedom of partons are thermalized, while the longitudinal motion is described by free streaming. When the energy density of the partonic system  drops to a certain critical value, the partons hadronize and the newly formed hadronic system freezes out. This sudden change is described with the help of the Landau matching conditions followed by the simulations done with THERMINATOR. The proposed scenario reproduces well the transverse-momentum spectra, the elliptic flow coefficient $v_2$, and the HBT radii of pions and kaons studied at RHIC (Au+Au collisions at \mbox{$\sqrt{s_{\rm NN}}$ = 200 GeV}). It also reproduces quite well the transverse-momentum spectra of hyperons. 
\end{abstract} 
%\submitto{\JPG}
%\maketitle 

\bigskip

The spacetime evolution of matter produced in the ultra-relativistic heavy-ion collisions is  described most successfully by the perfect-fluid hydrodynamics \mbox{\cite{Kolb:2003dz,Huovinen:2003fa,Shuryak:2004cy}}. With the initial conditions inferred from the Glauber model, the hydrodynamic approach describes very well the hadron transverse-momentum spectra and the elliptic flow coefficient $v_2$. With the Gaussian initial conditions, the hydrodynamic approach describes also the HBT radii \cite{Broniowski:2008vp}. 

Unfortunately, the use of the hydrodynamics is still challenged by the problem of early thermalization –-- the correct description of the data requires an early initialization time for hydrodynamics \cite{Bozek:2009ty}, which implicitly assumes a very early equilibration time. A possible solution to this problem was suggested in Ref.~\cite{Bialas:2007gn}, where the idea was proposed that only transverse degrees of freedom are thermalized, while the longitudinal motion is described by free streaming (a physical picture explained naturally in the string models \cite{Bialas:1999zg,Florkowski:2003mm}). In this framework, called below the {\it transverse hydrodynamics},  one can obtain the {\it parton} transverse-momentum spectra and $v_2$ which are consistent with the data \cite{Bialas:2007gn,Chojnacki:2007fi}~\footnote{We note that our formulation differs from the formalism of the transverse hydrodynamics introduced originally in \cite{Heinz:2002rs,Heinz:2002xf}. For a more detailed discussion of this issue see \cite{Bialas:2007gn}.}. In this paper we generalize this approach by inclusion of the sudden-hadronization transition modeled with the help of the Monte Carlo code THERMINATOR \cite{Kisiel:2005hn}. 

The equations of the transverse hydrodynamics follow from the energy-momentum conservation law, $\partial_\mu T_2^{\mu \nu}=0$, with the energy-momentum tensor \cite{Ryblewski:2008fx}
\begin{eqnarray}
T_2^{\mu \nu} = \frac{n_0}{\tau} \left[
\left(\varepsilon _2 + P_2\right) U^{\mu}U^{\nu} 
- P_2 \,\,\left( g^{\mu\nu} + V^{\mu}V^{\nu} \right)\,\, \right],
\label{tensorT1}
\end{eqnarray}
where $\tau = \sqrt{t^2 - z^2}$ is the longitudinal proper time and $n_0$ describes the density of transverse clusters in rapidity. The clusters are identified with the groups of partons having the same rapidity. They are two-dimensional (2D)  objects described by the 2D thermodynamic variables: $\varepsilon_2$, $P_2$, $s_2$ and $T_2$ (2D energy density, pressure, entropy density, and temperature, respectively). The 2D thermodynamic variables satisfy the standard thermodynamic identities: $\varepsilon_2 + P_2 = T_2 s_2$, $d\varepsilon_2 = T_2 ds_2$, and $dP_2 = s_2 dT_2$ (the baryon chemical potential is set equal to zero). The energy-momentum tensor (\ref{tensorT1}) includes also the two four-vectors,
\begin{eqnarray}
U^{\mu} &=& ( u_0 \cosh\eta,u_x,u_y, u_0 \sinh\eta), \quad V^{\mu} = (\sinh\eta,0,0,\cosh\eta),
\label{U}
\end{eqnarray}
where $u^\mu = \left(u^0, {\vec u}_\perp, 0 \right)$ is the hydrodynamic flow in the plane $z=0$, while \mbox{$\eta = 1/2 \ln ((t+z)/(t-z))$} is the spacetime rapidity. The four-vectors $U^\mu$ and $V^\mu$ satisfy the following normalization conditions: $U^\mu U_\mu = 1$, \mbox{$V^\mu V_\mu = -1$},  \mbox{$U^\mu V_\mu = 0$}. The four-vector $U^\mu$ corresponds to the flow four-velocity in the standard hydrodynamics. On the other hand, the term $V^\mu V^\nu$ in (\ref{tensorT1}) is responsible for vanishing of the longitudinal pressure, i.e., in the local rest-frame of the fluid element, where $U^\mu = (1,0,0,0)$ and $V^\mu = (0,0,0,1)$, we have
\begin{eqnarray}
\nonumber \\
T^{\mu \nu} = \frac{n_0}{\tau} \left(
\begin{array}{cccc}
\varepsilon _2 & 0 & 0 & 0 \\
0 & P_2 & 0 & 0 \\
0 & 0 & P_2 & 0 \\
0 & 0 & 0 & 0
\end{array} \right). \\ \nonumber
\end{eqnarray}
It should be emphasized that such a structure of the energy-momentum tensor appears in the theory of the {\it color glass condensate} and {\it glasma} for \mbox{$\tau \gg 1/Q_s$}, where \mbox{$Q_s \sim 1$ GeV} is the saturation scale expected at RHIC \cite{Kovchegov:2005ss,Krasnitz:2002mn}. This gives support for consideration of the transverse hydrodynamics as the appropriate description of the early evolution of matter. 

\bigskip
The equations of the transverse hydrodynamics are solved numerically with the equation of state $P_2 = \varepsilon_2/2 = \nu _g T_2^3 \zeta(3)/(2\pi)$, where $\nu_g = 16$ \cite{Ryblewski:2008fx}. Our analysis is restricted to the central rapidity region ($z \approx \eta \approx 0$), where the created system may be treated as boost-invariant.  The initial conditions assume that the 2D initial energy density at $\tau = \tau_{\rm i} = 1$ fm is proportional to the mixture of the wounded-nucleon density, $\rho_W \left({\vec x}_\perp \right)$, and the binary-collision density, $\rho_B \left({\vec x}_\perp \right)$,
\begin{equation}
\varepsilon_2\left(\tau_{\rm i},{\vec x}_\perp \right) 
= \frac{\nu_g \, T_{2}^{\,3} \left(\tau_{\rm i},{\vec x}_\perp \right) \zeta(3) }{\pi} 
\, \propto \,  \frac{1-\kappa}{2}\rho_W \left({\vec x}_\perp \right) + \kappa \rho_B \left({\vec x}_\perp \right),
\label{initcond}
\end{equation}	
where, following the PHOBOS studies of the centrality dependence of the hadron production \cite{Back:2004dy}, we take $\kappa=0.14$. The normalization constant required in (\ref{initcond}) determines the 2D initial central temperature of the system, $T_{2 \,\rm i}=T_2(\tau_{\rm i},0)$.

The Landau matching conditions have the form
\begin{equation}
T_2^{\mu \nu} U_\nu = T^{\mu \nu}_{3} U_\nu, \label{LMc1}
\end{equation} 
where $T^{\mu \nu}_{3}$ is the standard energy-momentum tensor of the perfect-fluid hydrodynamics
\begin{equation}
T_3^{\mu \nu} = (\varepsilon_3 + P_3) U^\mu U^\nu - P_3 g^{\mu \nu}. \label{LMc2}
\end{equation}  
Here $\varepsilon_3$ and $P_3$ are the three-dimensional (3D) energy density and pressure. Equations (\ref{LMc1}) and (\ref{LMc2}) give
\begin{equation}
\frac{n_0}{\tau} \varepsilon_2 = \varepsilon_3, \label{LMc3}
\end{equation} 
which should be supplemented by the requirement of the entropy growth,
\begin{equation}
\frac{n_0}{\tau} s_2 \leq s_3, \label{LMc4}
\end{equation} 	
where $s_3$ is the 3D entropy density. Dividing both sides of Eqs. (\ref{LMc3}) and (\ref{LMc4}) one obtains
\begin{equation}
T_2 \geq \frac{3 \,\varepsilon_3}{2 \,s_3}. \label{LMc5}
\end{equation} 	
In our calculation, $\varepsilon_3$, $P_3$, and $s_3$ are interpreted as the quantities characterizing the hadron resonance gas in equilibrium at the 3D temperature $T_{3 \, \rm f}$. This  means that the use of the Landau matching conditions is a formal scheme to describe the fast hadronization process; if the energy density of the system drops to $\varepsilon_3(T_{3\, \rm f})$, the transverse gluonic clusters change into a locally isotropic resonance gas. 

The considered model has three free parameters: $T_{2 \, \rm i}$, $T_{3 \, \rm f}$, and $n_0$ (the initial time $\tau_{\rm i}$ is always set equal to 1 fm). In the fitting procedure we first assume $n_0=1$ and try to adjust $T_{2 \, \rm i}$ and  $T_{3 \, \rm f}$ in such a way that a good description of the data is achieved. In this step, the energy-conservation condition (\ref{LMc3}) is used as the only constraint. In the next step we rescale the 2D temperature in such a way that the equality $T_2 = 3 \,\varepsilon_3 / (2 \,s_3)$ holds at the initial time $\tau=\tau_{\rm i}$. In this way we make certain that the condition (\ref{LMc5}) is fulfilled on the whole transition hypersurface. The rescaling of the temperature changes the energy density on the 2D side, however, this may be compensated by the appropriate change of $n_0$ (the flow profile is not affected by the rescaling of the temperature). In this way we are able to satisfy finally the conditions (\ref{LMc3}) and (\ref{LMc4}).

Our description of the hadronization process is quite radical. Yet, we add to it another radical assumption: the evolution of the isotropic (locally equilibrated) hadronic system is so short that the hadronization process may be followed immediately by freeze-out. Technically, this assumption is realized by the Monte-Carlo simulations done with THERMINATOR, where the 3D hadronization temperature $T_{3\,\rm f}$ is identified with the freeze-out temperature and the hadronization hypersurface becomes equivalent to the freeze-out hypersurface. 

Certainly, our method used to switch from the purely transverse hydrodynamic expansion to the locally isotropic system of hadrons and further to freeze-out is very much simplified. In particular, one expects that the isotropization process has a gradual character. More elaborate approaches would describe this kind of transformation using kinetic theory or dissipative hydrodynamics, for example, see \cite{Kovchegov:2005az,Bozek:2007di,Zhang:2008kj}. Alternatively, one may think of several direct developments of our scheme, for example, the transition criterion may be changed and/or the time evolution of the isotropic phase may be extended. Such developments are the subject of our current studies. Anyway, {\it in our opinion such more realistic smooth changes may be effectively replaced by the sudden but delayed processes}.
\begin{figure}[t]
\begin{center}
\includegraphics[angle=0,width=0.7\textwidth]{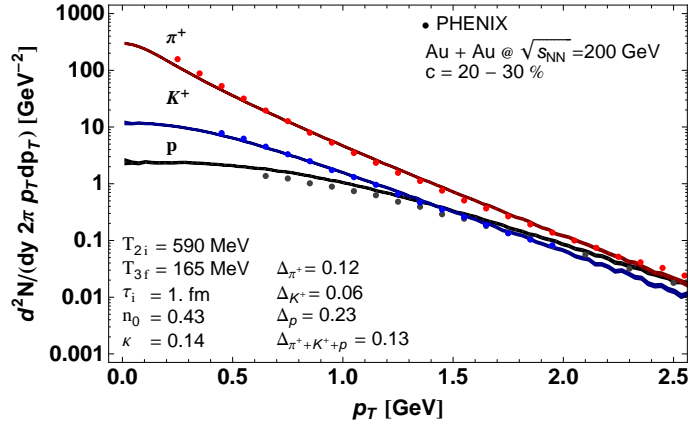}
\end{center}
\caption{\small Model transverse-momentum spectra of pions, kaons, and protons (solid lines) compared to the experimental PHENIX data (dots) \cite{Adler:2003cb}. }
\label{fig:figsppt-590-165}
\end{figure}	
\begin{figure}[t]
\begin{center}
\includegraphics[angle=0,width=0.65\textwidth]{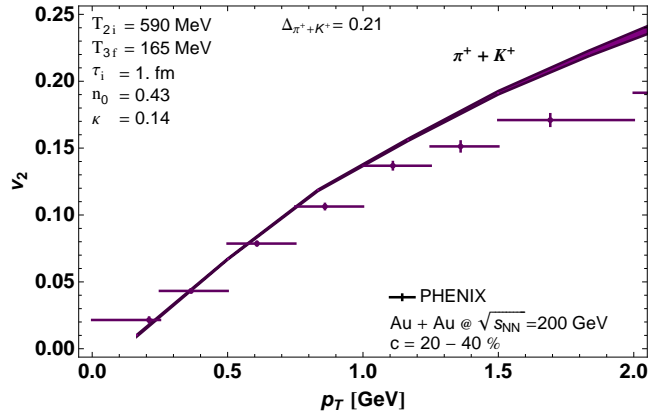}
\end{center}
\caption{\small Elliptic flow coefficient of pions and kaons (solid line) compared to the experimental PHENIX data (dashes) \cite{Adler:2003kt}. }
\label{fig:figv2pt-590-165}
\end{figure}	
\begin{figure}[t]
\begin{center}
\includegraphics[angle=0,width=0.7\textwidth]{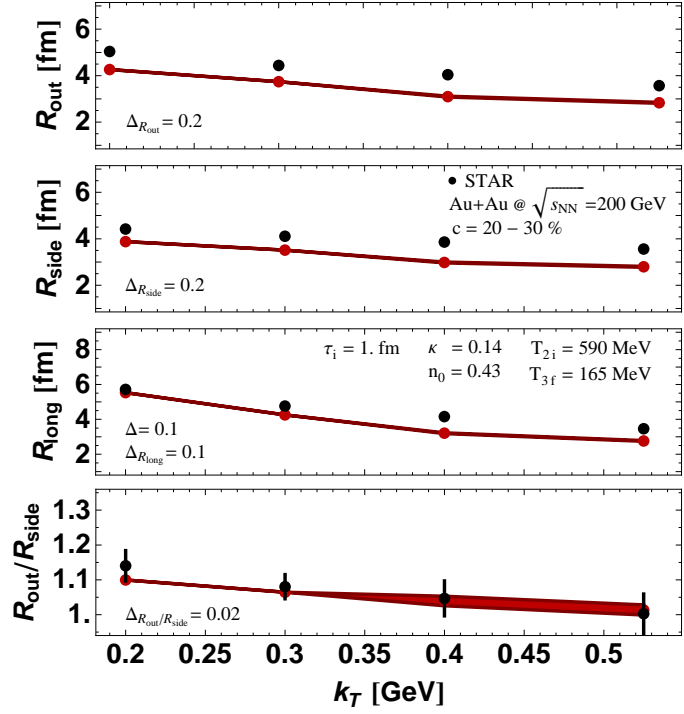}
\end{center}
\caption{\small Model results for the pionic HBT radii (solid lines) compared to the experimental STAR data  \cite{Adams:2004yc} (dots). The radii are plotted as the functions of the average transverse momentum of the pion pair $k_T$. The model parameters are the same as in Figs. \ref{fig:figsppt-590-165} and \ref{fig:figv2pt-590-165}. }
\label{fig:fighbt-590-165}
\end{figure}

\begin{figure}[t]
\begin{center}
\includegraphics[angle=0,width=0.75\textwidth]{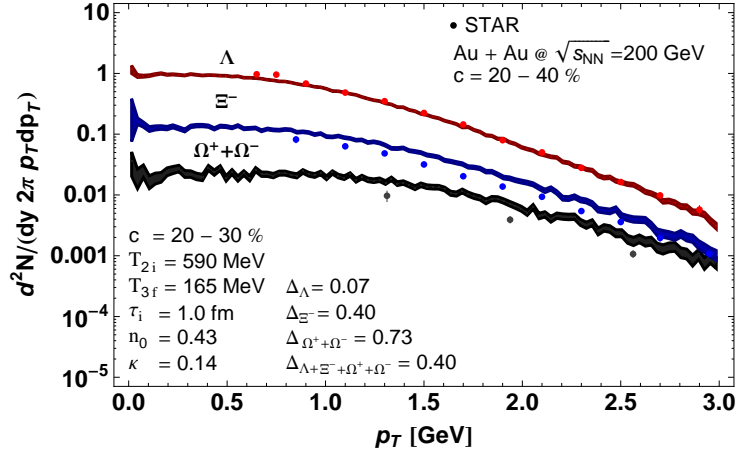}
\end{center}
\caption{\small Model results for hyperons compared to the experimental STAR data  \cite{Adams:2006ke} (dots). The model parameters are the same as in Figs. \ref{fig:figsppt-590-165} and \ref{fig:figv2pt-590-165}.}
\label{fig:figppts-590-165}
\end{figure}		

The results of our numerical calculations with the initial profiles corresponding to the non-central (\mbox{$c$ = 20-30\%}) Au+Au collisions at the beam energy \mbox{$\sqrt{s_{\rm NN}}$ = 200 GeV} are shown in Figs.~\ref{fig:figsppt-590-165} -- \ref{fig:figppts-590-165}. The values of the parameters used in the calculations  are: $T_{2 \, \rm i} =$ 590 MeV, $T_{3 \, \rm f} =$ 165 MeV, and $n_0=0.43$~\footnote{These parameters are slightly different from those used in \cite{Ryblewski:2009hm}, where we did not study the production of strangeness and the equation of state was of the form $P_2 = \varepsilon_2/2 = \nu _g T_2^3/(2\pi)$.} . We observe that our model reproduces well the transverse-momentum spectra of pions, kaons, and protons, Fig.~\ref{fig:figsppt-590-165}, and also the pion+kaon elliptic flow, Fig.~\ref{fig:figv2pt-590-165}, both measured in Au+Au collisions by PHENIX \cite{Adler:2003cb,Adler:2003kt}. The values of $\Delta$ indicate the average relative difference between the experimental data and the model results.

In Fig.~\ref{fig:fighbt-590-165} we show the model pionic HBT radii (solid lines) compared with the experimental STAR data (dots) \cite{Adams:2004yc}. The radii were calculated with the help of THERMINATOR which uses a two particle method with Coulomb corrections \cite{Kisiel:2006is}. The overall agreement with the data is quite good. The theoretical radii are by about 20\% smaller than the experimental values, but their $k_T$ dependence is very well reproduced. Moreover, the model ratio $R_{\rm out}/R_{\rm side}$ matches very well the experimental result. Slightly smaller theoretical values indicate that the real isotropization and freeze-out processes take longer time. This is, of course, an expected feature. 

The high freeze-out temperature obtained in our model suggests that one may be able to reproduce correctly the spectra of hyperons, as in the case of the single-freeze-out model \cite{Broniowski:2001uk}. The hyperons, due to their small interaction cross sections, are believed to decouple rather early at the temperature close to the typical chemical freeze-out temperature. Indeed, our calculations presented in Fig.~\ref{fig:figppts-590-165} show that the hyperon transverse-momentum spectra are reproduced reasonably well.

In the conclusions we state that the transverse-hydrodynamics model supplemented with the sudden isotropization transition and freeze-out is able to describe uniformly the soft pion and kaon data  (transverse-momentum spectra, $v_2$, and the HBT radii).  This brings further evidence that the assumption of the very fast 3D equilibration of matter produced in the relativistic heavy-ion collisions may be relaxed --- an earlier study showed that the RHIC data may be explained by the model where the standard hydrodynamic phase is preceded by the parton free streaming \cite{Broniowski:2008qk}. Our analysis of the HBT radii indicates, however, that the ultimate 3D equilibration is necessary to reproduce the experimentally measured value of $R_{\rm long}$.

\bigskip

%\bibliography{ref-rr}

\end{document}